\documentclass[letterpaper,conference]{IEEEtran}
\IEEEoverridecommandlockouts
\usepackage{amsmath, bm, amssymb, algpseudocode, algorithm, mathtools, svg,ifluatex, pgfplots, amsthm, caption, subcaption, tikz}
\usepackage{cite, bbm, accents}
\usepackage{graphicx}
\usepackage{textcomp}
\usepackage{xcolor}
\usepackage{stfloats,cuted}
\usepackage[acronym,shortcuts]{glossaries}
\usepackage[letterpaper, bottom=1.05in, top=0.75in, left=0.63in, right=0.63in]{geometry}
\usetikzlibrary{graphs, positioning, quotes, shapes.geometric, arrows}

\pgfplotsset{compat=1.15}

\newacronym{ba}{BA}{Blahut-Arimoto}
\newacronym{cc}{C-C}{capacity-cost}
\newacronym{cdc}{CDC}{capacity-distortion-cost}
\newacronym{wgd}{WGD}{Wasserstein gradient descent}
\newacronym{kld}{KLD}{Kullback-Leibler divergence}
\newacronym{rd}{R-D}{rate-distortion}
\newacronym{ac}{a.c.}{absolutely continuous}
\newacronym{rnd}{RND}{Radon-Nikodym derivative}
\newacronym{pdf}{PDF}{probability density function}
\newacronym{jko}{JKO}{Jordan-Kinderlehrer-Otto}
\newacronym{sgd}{SGD}{stochastic gradient descent}
\newacronym{nn}{NN}{neural network}
\newacronym{mci}{MCI}{Monte-Carlo integration}
\newacronym{is}{IS}{importance sampling}
\newacronym{mmse}{MMSE}{minimum mean square error}
\newacronym{csir}{CSIR}{channel state information at receiver}
\newacronym{isac}{ISAC}{integrated sensing and communications}
\newacronym{pm}{PM}{posterior mean}

\newcommand*{\brrr}[1]{\left \{ #1 \right \}}

\newcommand{\RR}{\mathbb{R}}

\newcommand{\calX}{\mathcal{X}}
\newcommand{\calY}{\mathcal{Y}}
\newcommand{\calZ}{\mathcal{Z}}
\newcommand{\calP}{\mathcal{P}}
\newcommand{\calW}{\mathcal{W}}

\newcommand{\calS}{\mathcal{S}}

\newcommand{\dx}{\mathrm{d}x}
\newcommand{\ds}{\mathrm{d}s}
\newcommand{\dy}{\mathrm{d}y}
\newcommand{\dz}{\mathrm{d}z}

\newcommand*{\st}{\mathrm{s. \, t. \,}}

\DeclareMathOperator*{\argmin}{arg\,min}
\DeclareMathOperator*{\arginf}{arg\,inf}

\theoremstyle{definition}

\tikzset{%
    smallblock/.style={draw, fill=white, minimum height=1em, minimum width=1em},
    block-common/.style={draw, fill=white, minimum height=1.5em, minimum width=4em},
    block/.style={rectangle, block-common, align=center},
    txtblock/.style={block, align=center, minimum height=4em},
    bigblock/.style={block, minimum height=8em},
    txtbigblock/.style={bigblock, align=center},
    input/.style={inner sep=1pt},       
    output/.style={inner sep=1pt},      
    sum/.style = {draw, fill=white, circle, minimum size=1.1em, inner sep=0pt,
      font={\small$+$}},
    prod/.style = {draw, fill=white, circle, minimum size=1.1em, inner sep=0pt,
      font={\normalsize$\times$}},
    pinstyle/.style = {pin edge={to-,thin,black}}
}

\begin{document}
\title{Computation of Capacity-Distortion-Cost Functions for Continuous Memoryless Channels
\thanks{X. Li, V. Andrei, U. M\"onich and H. Boche were supported in part by the
    German Federal Ministry of Education and Research (BMBF)
    in the programme “Souver\"an. Digital. Vernetzt.”
    within the research hub 6G-life under Grant 16KISK002.
    U. M{\"o}nich and H. Boche were also supported by the BMBF within the project "Post Shannon Communication - NewCom" under Grant 16KIS1003K.
    X. Li, V. Andrei, and H. Boche also acknowledge the financial support the financial support by the BMBF Projects QD-CamNetz, Grant
    16KISQ077, QuaPhySI, Grant 16KIS1598K, and QUIET, Grant 16KISQ093.
    }
}

\author{\IEEEauthorblockN{Xinyang Li\IEEEauthorrefmark{1},
Ziyou Tang,
Vlad C. Andrei\IEEEauthorrefmark{1}, Ullrich J. M{\"o}nich\IEEEauthorrefmark{1}, Fan Liu\IEEEauthorrefmark{2} and
Holger Boche\IEEEauthorrefmark{1}\IEEEauthorrefmark{3}}
\IEEEauthorblockA{\IEEEauthorrefmark{1}Chair of Theoretical Information Technology, Technical University of Munich, Munich, Germany\\
\IEEEauthorrefmark{2}School of Information Science and Engineering, Southeast University, Nanjing 210096, China\\
\IEEEauthorrefmark{1}BMBF Research Hub 6G-life,
\IEEEauthorrefmark{3}Munich Center for Quantum Science and Technology,
\IEEEauthorrefmark{3}Munich Quantum Valley\\
Email: \{xinyang.li, ziyou.tang, vlad.andrei, moenich\}@tum.de, fan.liu@seu.edu.cn, boche@tum.de}
}

\maketitle

\allowdisplaybreaks
\glsdisablehyper

\makeatletter
\def\ps@IEEEtitlepagestyle{%
  \def\@oddfoot{\mycopyrightnotice}%
  \def\@oddhead{\hbox{}\@IEEEheaderstyle\leftmark\hfil\thepage}\relax
  \def\@evenhead{\@IEEEheaderstyle\thepage\hfil\leftmark\hbox{}}\relax
  \def\@evenfoot{}%
}
\def\mycopyrightnotice{%
  \begin{minipage}{\textwidth}
  \centering \scriptsize
  Copyright~\copyright~20xx IEEE. Personal use of this material is permitted. Permission from IEEE must be obtained for all other uses, in any current or future media, including\\reprinting/republishing this material for advertising or promotional purposes, creating new collective works, for resale or redistribution to servers or lists, or reuse of any copyrighted component of this work in other works by sending a request to pubs-permissions@ieee.org.
  \end{minipage}
}
\makeatother

\begin{abstract}
This paper aims at computing the \ac{cdc} function for continuous memoryless channels, which is defined as the supremum of the mutual information between channel input and output, constrained by an input cost and an expected distortion of estimating channel state.
Solving the optimization problem is challenging because the input distribution does not lie in a finite-dimensional Euclidean space and the optimal estimation function has no closed form in general. We propose to adopt the Wasserstein proximal point method and parametric models such as \acp{nn} to update the input distribution and estimation function alternately. To implement it in practice, the \ac{is} technique is used to calculate integrals numerically, and the Wasserstein gradient descent is approximated by pushing forward particles. The algorithm is then applied to an \ac{isac} system, validating theoretical results at minimum and maximum distortion as well as the random-deterministic trade-off.
% Numerical results demonstrate the high accuracy of the proposed approach.
\end{abstract}


\glsresetall

\section{Introduction}\label{sec:intro}
Advanced wireless networks are envisioned to incorporate \ac{isac} functionalities\cite{liu2022integrated}, enabling the communication systems to simultaneously transmit data and gather environmental information. From the perspective of information theory, \ac{isac} is typically modeled as a state-dependent channel\cite{liu2022survey,ahmadipour2022information,li2025simultaneousdecodingapproachjoint}, where the goal is to jointly convey messages reliably and estimate channel states accurately. Compared to traditional systems with only communication capabilities, whose performance limits are assessed by capacity-cost functions\cite{gallager1968information}, \ac{isac} systems are commonly characterized by \ac{cdc} functions\cite{zhang2011joint,ahmadipour2022information} that describe the trade-off between maximum data rate, state estimation distortion, and input cost.

Computing \ac{cdc} functions helps in understanding the optimal input distributions and state estimation functions at different operating points, thereby designing suitable waveforms and estimators in practice to achieve better performance. For discrete memoryless channels with finite alphabets of the input, output and channel state, \cite{ahmadipour2022information} proposes a Blahut-Arimoto-type \cite{blahut1972computation,arimoto1972algorithm} method to compute the \ac{cdc} function numerically, where the input distribution is updated iteratively, and the state estimator searches globally from the finite reconstruction set.

As most physical systems involve continuous-valued signals and states, extending the computation of \ac{cdc} functions to continuous alphabets becomes crucial for practical applications. However, such an extension is difficult and faces two problems. Firstly, the input distributions of continuous alphabets cannot be represented as finite-length vectors within an Euclidean subspace, making standard optimization tools for finite-dimensional spaces inapplicable. Secondly, the global search method used for the optimal estimation becomes infeasible for the continuous case, and the analytic form of the optimal estimator is unknown in general. To the best of our knowledge, besides the related works\cite{chang1988calculating,dauwels2005numerical,Li2506:Computing} attempting to compute capacity-cost functions for continuous channels, the aforementioned problems have not been fully analyzed before.

To tackle these challenges, we first adopt the optimization framework in the Wasserstein space\cite{ambrosio2008gradient,santambrogio2017euclidean}.
% which contains all probability measures with finite second moments and is equipped with the Wasserstein distance as the distance measure. 
We reformulate the optimization problem of the \ac{cdc} function over input distribution as a Wasserstein proximal point problem, following our recent work\cite{Li2506:Computing}, and establish a gradient descent type method in the Wasserstein space. As for the state estimator, we propose to parameterize it as a differentiable function, such as \ac{nn}, if its optimal form has no analytic expression. The overall algorithm updates the input distribution and estimator parameters alternately until convergence. Numerically, the Wasserstein gradient descent is implemented by transporting a set of particles in the input space by a sequence of pushforward operators. Moreover, the involved integrals are computed approximately by the \ac{is} technique. Simulation results are provided for a simple monostatic-downlink \ac{isac} system, showing a convincing performance of the proposed method.

\section{Background}
\subsection{Capacity-Distortion-Cost Function}

We shall consider the \ac{cdc} function studied in\cite{ahmadipour2022information} for continuous memoryless channels. Let $\calX, \calY, \calZ, \calS$ be compact subsets of Euclidean spaces. The memoryless channel is assumed to be characterized by a conditional \ac{pdf} $p_{YZ|XS}$ mapping the input $X\in \calX$ to two outputs $(Y,Z)\in \calY\times \calZ$ and controlled by the channel state $S\in\calS$, which is generated from a \ac{pdf} $p_S$. Such a channel is said to be continuous if $p_{YZ|XS}$ is a continuous function in $X$ and $S$. Define the input cost function $b: \calX \to [0, \infty)$ and the state estimation distortion function $d: \calS \times \hat{\calS} \to [0,\infty)$ with the state reconstruction space $\hat{\calS}$. We further assume that $b$ and $d$ are differentiable functions and  $p_{YZ|XS}$ is differentiable with respect to $X$ in order to adopt the proposed method. With the same setup\footnote{In\cite{ahmadipour2022information}, the communication receiver is also assumed to know the channel state $S$. This makes no difference in analysis if one regards $S$ as another channel output with $Y$.} as\cite{ahmadipour2022information}, suppose that the communication receiver decodes the message reliably upon receiving the channel output $Y$, while a state estimator wishes to estimate $S$ based on $X$ and the other output $Z$, the maximum achievable communication data rate, under the expected input cost constraint $B$ and expected estimation distortion constraint $D$, is given by the \ac{cdc} function
\begin{equation}\label{eq:cdb}
\begin{split}
    &C(D,B) \triangleq \sup I(X; Y)\\
    &\hspace{10mm}= \sup \int_{\calX} \mu(\dx) \int_{\calY} p_{Y|X}(y|x) \log \frac{p_{Y|X}(y|x)}{P_Y(y)} \dy\\
    & \st  \int_{\calX} b(x) \mu(\dx) \le B, \\
    & \int_{\calX} \mu(\dx) \int_{\calS\times \calZ} p_S(s) p_{Z|XS}(z|x,s) d(s, h^*(x,z)) \ds \dz \le D,
\end{split}
\end{equation}
where the optimal estimation function $h^*: \calX\times \calZ \to \hat{\calS}$ is
\begin{equation}\label{eq:optest}
    h^*(x,z) \triangleq \argmin_{s'\in\hat{\calS}} \int_{\calS} p_{S|XZ}(s|x,z) d(s,s') \ds
\end{equation}
and
\begin{align}
    p_{Y|X}(y|x) &= \int_{\calS\times\calZ} p_{YZ|XS}(y,z|x,s) p_S(s) \ds \dz,\\
    p_{Y}(y) &= \int_{\calX} \mu(\dx) p_{Y|X}(y|x),\\
    p_{Z|XS}(z|x,s) &= \int_{\calY} p_{YZ|XS}(y,z|x,s) \dy,\\
    p_{S|XZ}(s|x,z) &= \frac{p_S(s) p_{Z|XS}(z|x,s)}{\int_{\calS} p_S(s) p_{Z|XS}(z|x,s) \ds}\label{eq:p_s__xz}.
\end{align}
The supremum is taken over all $\mu \in \calP(\calX)$ with $\calP(\calX)$ the set of all the Borel probability measures on $\calX$\cite{billingsley2012probability}. Although the achievability and converse for \eqref{eq:cdb} are only proved for the discrete memoryless channel in\cite{ahmadipour2022information}, the extension to the continuous case follows the similar arguments as in\cite{gallager1968information,yeung2008information}.

It is noted that $C(D,B)$ is not restricted to describing the fundamental limits of scenario in\cite{ahmadipour2022information} where $Z$ is the generalized feedback and the state is estimated at the transmitter side. By setting $Y=Z$ and letting the receiver jointly recover the message and channel state, $C(D,B)$ becomes the \ac{cdc} function derived in\cite{zhang2011joint}. Similarly, the performance of the bistatic-downlink \ac{isac} system is also characterized by $C(D,B)$, treating $Y$ and $Z$ as the output of downlink and bistatic sensing link, respectively, as shown in \cite{li2025simultaneousdecodingapproachjoint}.

In\cite{ahmadipour2022information}, a Blahut-Arimoto type algorithm\cite{blahut1972computation,arimoto1972algorithm} is proposed to solve the optimization problem in $C(D,B)$ for the discrete case. However, optimizing over $\calP(\calX)$ for the continuous case is not straightforward since it can not be treated as a finite-dimensional vector subspace. Moreover, the $\argmin$ problem in \eqref{eq:optest} poses another challenge if $\hat{\calS}$ is not discrete, from which finding the optimum is infeasible unless one can derive the analytic expression of $h^*$. This motivates us to utilize more advanced optimization techniques, including the Wasserstein proximal point method and parametric models.

\subsection{Wasserstein Space}

Given a Borel space $(\calX, \mathcal{B}(\calX))$ with $\calX$ a compact subset of the Euclidean space, the set of all Borel probability measures with finite second moments is denoted as
\begin{equation*}
    \calP_2(\calX) \triangleq \brrr{\mu \in  \calP(\calX) \Big| \int_{\calX} \| x \|_2^2\  \mu(\dx) < + \infty }.
\end{equation*}
Let $T: \calX \to \calX$ be a $\mathcal{B}(\calX)$-measurable function, the pushforward measure\cite{ambrosio2008gradient} of $\mu$ under $T$ is $T_\#\mu$ such that
\begin{equation}
    T_\#\mu(B) = \mu(T^{-1}(B)), \quad \forall B \in \mathcal{B}(\calX).
\end{equation}
The 2-Wasserstein distance\cite{ambrosio2008gradient} between two measures $\mu, \nu \in \calP_2(\calX)$ is defined as
\begin{equation}\label{eq:w2prob}
     W_2(\mu, \nu) \triangleq \left (\min_{\pi \in \Pi(\mu, \nu)}\int_{\calX\times \calX} \|x - y \|_2^2\  \pi(\dx,\dy) \right)^{\frac{1}{2}},
\end{equation}
where $\Pi(\mu, \nu)$ is the set of all couplings between $\mu$ and $\nu$, i.e., all the product probability measures on $\calX \times \calX$ such that the first and second marginals are $\mu$ and $\nu$ respectively. The optimal coupling $\pi^*$ achieving the minimum in \eqref{eq:w2prob} is called the optimal transport plan. In the following, we call $W_2$ the Wasserstein distance for convenience. The optimization problem involved in $W_2$ can be recast as the dual form
\begin{equation}
    \sup_{\varphi \in L^1(\mu)} \int \varphi(x) \mu(\dx) + \int \varphi^c(x) \nu(\dx)
    \label{eq:wassersteindual}
\end{equation}
where $\varphi^c$ is the c-transform of $\varphi$:
\begin{equation}
    \varphi^c(y) = \sup_{x \in \RR^n} \varphi(x) - \| x- y\|_2^2, \quad \forall y \in \RR^n,
\end{equation}
and the resulting optimizer $\varphi$ to \eqref{eq:wassersteindual} is called the Kantorovich potential.
Due to Brenier's theorem\cite{brenier1991polar, santambrogio2017euclidean}, if $\mu$ is \ac{ac}, the optimal transport plan $\pi^*$ is uniquely determined by $\pi^* = (\mathrm{Id}, T_\mu^\nu)_\# \mu$ with the identity map $\mathrm{Id}$ on $\calX$, the optimal transport map $T_\mu^\nu(x) = x - \nabla_x \varphi(x)$ and the Kantorovich potential $\varphi$.

It is shown that $W_2$ satisfies the properties of a metric, and thus $\calW_2 \triangleq (\calP_2(\calX), W_2)$ forms a metric space called the Wasserstein space\cite{villani2009optimal}. It also turns out that $\calW_2$ is equipped with a differential structure: for a functional $F: \calP_2(\calX) \to \RR$, its first variation\cite{santambrogio2015optimal}, if exists, is given by $\frac{\delta F}{\delta \mu}: \calX \to \RR$ such that for all $\chi$ with $\mu + \epsilon \chi \in \calP_2(\calX) $
\begin{equation*}
    \lim_{\epsilon \to 0^+} \frac{F(\mu + \epsilon \chi) - F(\mu)}{\epsilon} = \int_{\calX} \frac{\delta F}{\delta \mu} (x)\  \chi(\dx).
\end{equation*}
A simple example is when $F(\mu) = \int V(x) \mu(\dx)$ for a function $V: \calX \to \RR$, then $\frac{\delta F}{\delta \mu} (x) = V(x)$. Moreover, the first variation of the function $\mu \mapsto W_2^2(\mu, \nu)$ is given by $\frac{\delta W_2^2(\cdot, \nu)}{\delta \mu} (x) = \varphi(x)$ up to additive constants with the Kantorovich potential $\varphi$\cite{santambrogio2015optimal}.

\subsection{Importance Sampling}

Throughout this paper, we shall frequently encounter the computation of integrals, such as \eqref{eq:cdb}-\eqref{eq:p_s__xz}. If there is no analytic form, numerical methods can be performed. Common approaches include the finite-partition method and \ac{is}, where the former can yield accurate results but suffer from exponentially increasing computational complexity in the number and dimension of random variables. The \ac{is} technique, on the other hand, can balance complexity and accuracy by sampling different numbers of particles.

To compute an integral $I = \int f(a) \mathrm{d} a$, the basic idea of \ac{is} is to draw a set of $N$ samples $\{a_i\}_{i=1}^{N}$ from a pre-defined distribution $q(a)$, called the importance distribution\cite{candy2016bayesian}, and then approximate $I$ by 
\begin{equation*}
    I \approx \hat{I} = \int  \frac{1}{N} \sum_{i=1}^{N} \frac{f(a)}{q(a)} \delta_{a_i}(\mathrm{d}a) = \frac{1}{N} \sum_{i=1}^{N} \frac{f(a_i)}{q(a_i)},
\end{equation*}
where $\delta_{a_i}$ is the Dirac measure at $a_i$. As an example, to compute the integral $\int_{\calS} p_S(s) p_{Z|XS}(z|x,s) \ds$ in \eqref{eq:p_s__xz}, we set $p_S$ as the importance distribution for simplicity if possible\footnote{For the sake of convenience, we shall assume one can sample particles directly from $p_S$ and $p_{YZ|XS}$ conditioned on any $(x,s)$ in this paper.} and sample a set of $\{s_i\}_{i=1}^{N}$ from it. Then, $p_{S|XZ}(s|x,z)$ is approximated as
\begin{equation}\label{eq:p_s__xz_is}
    \hat{p}_{S|XZ}(s|x,z) = \frac{p_S(s) p_{Z|XS}(z|x,s)}{\frac{1}{N}\sum_{i=1}^N p_{Z|XS}(z|x,s_i)}.
\end{equation}
As a result, the integral in \eqref{eq:optest} can be numerically evaluated as
\begin{equation}\label{eq:p_s__xz2}
\begin{split}
    \int_{\calS} p_{S|XZ}(s|x,z) d(s,s') \approx&\frac{1}{N} \sum_{j=1}^N \frac{\hat{p}_{S|XZ}(s_j|x,z) d(s_j, s')}{p_S(s_j)}\\
    =& \frac{1}{N} \sum_{j=1}^N \frac{p_{Z|XS}(z|x,s_j)d(s_j, s')}{\frac{1}{N}\sum_{i=1}^N p_{Z|XS}(z|x,s_i)}.
\end{split}
\end{equation}

\section{Proposed Method}
\subsection{Problem Formulation}

Since the analytic expression of $h^*$ in \eqref{eq:optest} is unknown in general, we propose to parameterize the estimation functions as $h_\theta : \calX\times \calZ \to \hat{\calS}$ with $\theta$ from a parameter space $\Theta$. Common parameterization approaches include models like linear regression or \acp{nn}. The optimization over $\theta\in \Theta$ is then involved in the overall problem. Moreover, we require that $\mu \in \calP_2(\calX)$, which is not restrictive in reality. Such a constraint is satisfied automatically if one sets $b(x) = \|x\|_2^2$ and $B < \infty$. Similarly to the discrete case, the dual variables $\lambda \ge 0$ and $\beta \ge 0$ are introduced, and \eqref{eq:cdb} is reformulated as
\begin{equation}\label{eq:optlag}
    \inf_{\mu \in \calP_2(\calX), \theta\in\Theta} L_{\lambda,\beta}(\mu, \theta)
\end{equation}
with the corresponding Lagrangian function
\begin{equation}\label{eq:lagrang}
\begin{split}
    &L_{\lambda,\beta}(\mu, \theta) \triangleq \int_{\calX}\mu(\dx)\\
    &\hspace{5mm}\cdot\left( \lambda b(x) + \beta \int_{\calS\times \calZ} p_S(s) p_{Z|XS}(z|x,s) d(s, h_\theta(x,z)) \ds \dz\right.\\
    &\hspace{15mm}\left.- \int_{\calY} p_{Y|X}(y|x) \log \frac{p_{Y|X}(y|x)}{P_Y(y)} \dy\right).
\end{split}
\end{equation}

Optimization over $\mu$ and $\theta$ in \eqref{eq:optlag} can be conducted alternately. The update of $\mu$ involves the proximal point method in Wasserstein space, while updating $\theta$ depends on the chosen parametric model.

\subsection{Update of $\mu$}

We shall follow the ideas in \cite{Li2506:Computing}, which extends the proximal point reformulation of the Blahut-Arimoto algorithm \cite{matz2004information, naja2009geometrical} to the Wasserstein space. By fixing $\theta$, the update of $\mu$ at step $t$ requires solving a Wasserstein proximal point problem
\begin{equation}\label{eq:wasser_prox}
    \mu^{(t)} = \arginf_{\mu\in \calP_2(\calX)} L_{\lambda,\beta} (\mu,\theta)- D(p_Y \| p_Y^{(t-1)}) + \frac{W_2^2(\mu, \mu^{(t-1)})}{\tau_t}.
\end{equation}
The optimality condition for the $t$-th step problem is that the first variation of its objective function with respect to $\mu$ is equal to a constant value\cite{ambrosio2008gradient}. Note that
\begin{equation}
    L_{\lambda,\beta} (\mu,\theta) - D(p_Y \| p_Y^{(t-1)}) = \int_{\calX}\mu(\dx) V_{\lambda,\beta}^{(t)}(x,\theta)
\end{equation}
with
\begin{equation}
\begin{split}
   &V_{\lambda,\beta}^{(t)}(x,\theta) \triangleq\\
   &\lambda b(x) + \beta \int_{\calS\times \calZ} p_S(s) p_{Z|XS}(z|x,s) d(s, h_\theta(x,z)) \ds \dz\\
    &\hspace{15mm}- \int_{\calY} p_{Y|X}(y|x) \log \frac{p_{Y|X}(y|x)}{P^{(t-1)}_Y(y)} \dy.
\end{split}
    \label{eq:Vx}
\end{equation}
This leads to 
\begin{equation*}
    V_{\lambda,\beta}^{(t)}(x,\theta) + \frac{\varphi^{(t)}(x)}{\tau_t} = \mathrm{const},
\end{equation*}
with the Kantorovich potential $\varphi^{(t)}(x)$ associated with $W_2(\mu, \mu^{(t-1)})$,
and thus
\begin{equation*}
    \nabla_x V_{\lambda,\beta}^{(t)}(x,\theta) + \frac{\nabla_x \varphi^{(t)}(x)}{\tau_t} = 0.
\end{equation*}
Leveraging Brenier's theorem, the optimal transport plan from $\mu^{(t)}$ to $\mu^{(t-1)}$ is obtained as
\begin{equation}
    T_{\mu^{(t)}}^{\mu^{(t-1)}}(x) \triangleq x - \nabla_x \varphi^{(t)}(x) = x + \tau_t \nabla_x  V_{\lambda,\beta}^{(t)}(x,\theta),
    \label{eq:muk_to_muk-1}
\end{equation}
and for sufficiently small $\tau_t$, the optimal transport plan from $\mu^{(t-1)}$ to $\mu^{(t)}$ is approximated as
\begin{equation}
    T^{(t)}_\theta(x) \triangleq x - \tau_t \nabla_x  V_{\lambda,\beta}^{(t)}(x,\theta).
    \label{eq:update_rule_x}
\end{equation}
Therefore, the update rule for $\mu^{(k)}$ becomes
\begin{equation}
    \mu^{(t)} = (T_{\mu^{(t)}}^{\mu^{(t-1)}})^{-1}_\# \mu^{(t-1)} \approx T^{(t)}_{\theta \#} \mu^{(t-1)}.
    \label{eq:update_rule_mu}
\end{equation}

% It is worth mentioning that the approximated solution \eqref{eq:update_rule_mu} to \eqref{eq:wasser_prox} can be considered analogously to the explicit Euler scheme of the gradient flow in Euclidean spaces\cite{santambrogio2017euclidean}, 
% which is computationally efficient but may suffer from potential instability problems. An implicit scheme is also possible to derive but is beyond the scope of this paper. Nevertheless, simulation results show that the explicit scheme can still yield remarkable performance.

The implementation of \eqref{eq:update_rule_mu} in practice is commonly realized by the particle method\cite{wibisono2018sampling,yang2024estimating}. Given $\mu \in \calP_2(\calX)$ and a set of $N$ particles $\{x_i\}_{i=1}^N$ sampled from $\mu$, the empirical probability measure $\mu_N \triangleq \frac{1}{N} \sum_{i=1}^N \delta_{x_i}$,
represented by averaging Dirac measures at sample points, is usually used to approximate $\mu$.
% where $\delta_{x_i}$ is the Dirac measure at $x_i$ defined as for any $A \in \mathcal{B}(\calX)$
% \begin{equation*}
%     \delta_{x_i}(A) \triangleq \begin{cases}
%     1, & x_i \in A,\\
%     0, & x_i \notin A.
%     \end{cases}
% \end{equation*}
It also shows that the pushforward measure $T_\# \mu_N$ for any measurable function $T$ is again an empirical measure consisting of a set of particles $\{y_i\}_{i=1}^N$ with $y_i = T(x_i)$ for all $i$.
Hence, starting from a initial empirical measure $\mu_N^{(0)} = \frac{1}{N} \sum_{i=1}^N \delta_{x_i^{(0)}}$, the update rule \eqref{eq:update_rule_mu} results in a sequence of particle sets $\{x_i^{(t)}\}_{i=1}^N$ for $t=1,2,...$ with
\begin{equation}
    x_i^{(t)} = T_\theta^{(t)}(x_i^{(t-1)}) = x_i^{(t-1)} - \tau_t \nabla_x V_{\lambda,\beta}^{(t)}(x_i^{(t-1)}, \theta), \  \forall i.
    \label{eq:update_particle}
\end{equation}

Accordingly, the numerical computation of $V_{\lambda,\beta}^{(t)}(x, \theta)$ and $\nabla_x V_{\lambda,\beta}^{(t)}(x, \theta)$ can be conducted by \ac{is} technique and automatic differentiation, denoted as $\hat{V}_{\lambda,\beta}^{(t)}(x, \theta)$ and $\widehat{\nabla_x V}_{\lambda,\beta}^{(t)}(x, \theta)$, which is similar to \cite{Li2506:Computing} and thus is omitted here.
% where $p_Y^{(t-1)}(y)$ in $V_{\lambda,\beta}^{(t)}(x_i^{(t-1)}, \theta)$ is computed based on $\mu_N^{(t-1)}$ as
% \begin{equation}
% \begin{split}
%     p_Y^{(t-1)}(y) &= \int_{\calX} \frac{1}{N} \sum_{i=1}^N p_{Y|X}(y|x) \delta_{x^{(t-1)}_i}(\dx)\\
%     &= \frac{1}{N} \sum_{i=1}^N  p_{Y|X}(y|x^{(t-1)}_i).
%     \label{eq:py_approx}
% \end{split}
% \end{equation}
% By applying \ac{is}, we can evaluate $V_{\lambda,\beta}^{(t)}(x, \theta)$ and $\nabla_x V_{\lambda,\beta}^{(t)}(x, \theta)$ as
% \begin{align}
%     \hat{V}_{\lambda,\beta}^{(t)}(x, \theta) &= 
% \end{align}

\subsection{Update of $\theta$}

% If the explicit form of $h^*$ is known, it suffices to update $\mu$ at each step. As a special case,
% we consider the quadratic distortion function $d(s, \hat{s}) = (s-\hat{s})^2$, for which the optimal estimation function is the \ac{mmse} estimator and takes the form of the posterior mean
% \begin{equation}\label{eq:h_Q}
%     h^*_Q(x,z) = \int_{\calS} s\, p_{S|XZ}(s|x,z) \ds,
% \end{equation}
% which can be evaluated numerically with \ac{is} as in \eqref{eq:p_s__xz2}.

As an example of parametric methods, we shall adopt the \ac{nn} model in this paper, and updating $\theta$ corresponds to training the network to reduce the loss function
\begin{equation}\label{eq:nnloss}
    \int_{\calX}\mu(\dx)\int_{\calS\times \calZ} p_S(s) p_{Z|XS}(z|x,s) d(s, h_\theta(x,z)) \ds \dz.
\end{equation}
With the particle method and \ac{is} technique, the empirical loss value can be computed, based on which
the \ac{sgd} is performed to update $\theta$.

It is noted that the particle update step \eqref{eq:update_particle} involves the computation of $\nabla_x V_{\lambda,\beta}^{(t)}$, and in turn requires $h_\theta$ to be differentiable with respect to $x$ (c.f. \eqref{eq:Vx}), which is equivalent to computing the gradient of the \ac{nn} output to input.

\subsection{Algorithm}

Combining the alternating update rules for $\mu$ and $\theta$, we obtain the algorithm for numerical computation of $C(D,B)$ and summarize it in Algorithm \ref{alg:compute}. In addition, the dual ascent method\cite{boyd2011distributed} can be applied to update the dual variables $\lambda$ and $\beta$ at each step if one wishes to compute $C(D,B)$ at certain values of $D$ and $B$.

Similarly to the convergence analysis in\cite{Li2506:Computing}, the local convergence of \eqref{eq:wasser_prox} and \eqref{eq:update_rule_mu} in the Wasserstein space is guaranteed if certain conditions on $\tau_t$ are satisfied and the estimation function is fixed throughout. The convergence behavior involving the alternating update of $\theta^{(t)}$ highly depends on the chosen parametric model $h_\theta$, which is thus left as our future work.

\begin{algorithm}[t]
\caption{Numerical computation of \ac{cdc} function}
\begin{algorithmic}
    \State \textbf{Inputs:} 
    \State Channel distribution $p_{YZ|XS}$ with marginals $p_{Y|XS}$ and $p_{Z|XS}$; channel state distribution $p_S$; cost function $b$; distortion function $d$; initial measure $\mu^{(0)}$; number of particles $N$; number of \ac{is} samples $N_s$, $N_y$, $N_z$ for $S, Y, Z$; sequence of step sizes $\tau_1, \tau_2,...$; \ac{nn} learning rate $\epsilon_1, \epsilon_2, ...$; initial dual multipliers $\lambda_0$, $\beta_0$
    \State \textit{Optional}: cost upper bound $B$; distortion upper bound $D$; sequences of step sizes for dual ascent update $\alpha_0, \alpha_1,...$ and $\gamma_0, \gamma_1,...$
    \State \textbf{Initialize:}
    \State Sample a particle set $\{x_i^{(0)}\}_{i=1}^N$ from $\mu^{(0)}$;
    \State Randomly initialize \ac{nn} parameters $\theta^{(0)}$;
    \State $t \gets 0$
    \While{convergence condition not met}
    \State Sample $\{s_{j}^{(t)}\}_{j=1}^{N_s}$ from $p_{S}$
    \State Sample $\{y_{k_{i,j}}^{(t)}\}_{k_{i,j}=1}^{N_y}$ from $p_{Y|XS}(y|x_i^{(t)}, s_{j}^{(t)})\  \forall i,j$
    \State Sample $\{z_{l_{i,j}}^{(t)}\}_{l_{i,j}=1}^{N_z}$ from $p_{Z|XS}(z|x_i^{(t)}, s_{j}^{(t)})\  \forall i,j$
    \State $\hat{p}^{(t)}_{Y|X}(y_{k_{i,j}}^{(t)}|x_i^{(t)}) \gets \frac{1}{N_s}\sum_{j'=1}^{N_s} p_{Y|XS}(y_{k_{i,j}}^{(t)}|x_i^{(t)}, s_{j'}^{(t)})$
    \State $\hat{p}^{(t)}_{Y}(y_{k_{i,j}}^{(t)}) \gets \frac{1}{NN_s}\sum_{i'=1}^N\sum_{j'=1}^{N_s} p_{Y|XS}(y_{k_{i,j}}^{(t)}|x_{i'}^{(t)}, s_{j'}^{(t)})$
    \State Compute $\hat{V}_{\lambda,\beta}^{(t+1)}(x_i^{(t)}, \theta^{(t)})$ and $\widehat{\nabla_x V}_{\lambda,\beta}^{(t+1)}(x_i^{(t)}, \theta^{(t)})$
    % \begin{equation*}
    % \begin{split}
    %     &\hspace{-7mm}\hat{V}_{\lambda,\beta}^{(t+1)}(x_i^{(t)}, \theta^{(t)})\gets \lambda b(x_i^{(t)}) \\
    %     &+ \frac{1}{N_sN_z}\sum_{j=1}^{N_s}\sum_{l_{i,j}=1}^{N_z} \beta d(s_j, h_{\theta^{(t)}}(x_i^{(t)}, z_{l_{i,j}}^{(t)}))\\
    %     &- \frac{1}{N_sN_y}\sum_{j=1}^{N_s}\sum_{k_{i,j}=1}^{N_y} \log \frac{\hat{p}^{(t)}_{Y|X}(y_{k_{i,j}}^{(t)}|x_i^{(t)})}{\hat{p}^{(t)}_{Y}(y_{k_{i,j}}^{(t)})}
    % \end{split}
    % \end{equation*}
    \State $\hat{L}^{(t)} \gets \frac{1}{N} \sum_{i=1}^N \hat{V}_{\lambda,\beta}^{(t+1)}(x_i^{(t)}, \theta^{(t)})$ \Comment{see \eqref{eq:lagrang}}
    \State $\hat{B}^{(t)} \gets \frac{1}{N} \sum_{i=1}^N b(x_i^{(t)})$
    \State $\hat{D}^{(t)} \gets \frac{1}{NN_sN_z}\sum_{i=1, j=1}^{N, N_s} \sum_{l_{i,j}=1}^{N_z} d(s_j, h_{\theta^{(t)}}(x_i^{(t)}, z_{l_{i,j}}^{(t)}))$
    \State \Comment{see \eqref{eq:nnloss}}
    \State $\hat{R}^{(t)} \gets  \lambda_t \hat{B}^{(t)} + \beta_t\hat{D}^{(t)} - \hat{L}^{(t)}$
    \State $x_i^{(t+1)} \gets x_i^{(t)} - \tau_{t+1} \widehat{\nabla_x V}_{\lambda,\beta}^{(t+1)}(x_i^{(t)}, \theta^{(t)})$ \Comment{see \eqref{eq:update_particle}}
    \State $\theta^{(t+1)} \gets \theta^{(t)} - \epsilon_{t+1} \nabla_\theta \hat{D}^{(t)}$
    % \State Update $\theta^{(t+1)}$ via \ac{sgd} based on $\hat{D}^{(t)}$
    \If{dual ascent update required}
    \State $\lambda_{t+1} \gets \max\{0,\lambda_t + \alpha_t (\hat{B}^{(t)} - B)\}$
    \State $\beta_{t+1} \gets \max\{0,\beta_t + \gamma_t (\hat{D}^{(t)} - D)\}$ 
    \Else
    \State $\lambda_{t+1} \gets \lambda_t$
    \State $\gamma_{t+1} \gets \gamma_t$
    \EndIf
    \State $t \gets t+1$
    \EndWhile
    \State Evaluate $\hat{L}^{(t)}$, $\hat{R}^{(t)}$, $\hat{B}^{(t)}$, $\hat{D}^{(t)}$ on $\{x_i^{(t)}\}_{i=1}^N$
    \State \textbf{Return:}
    \State $\{x_i^{(t)}\}_{i=1}^N$, $\hat{L}^{(t)}$, $\hat{R}^{(t)}$, $\hat{B}^{(t)}$, $\hat{D}^{(t)}$
\end{algorithmic}\label{alg:compute}
\end{algorithm}

\section{Numerical Results}
We consider a monostatic-downlink \ac{isac} model defined as
\begin{equation}
\begin{aligned}
    y &= x + n_1, \\
    z &= a(\phi)x + n_2,
\end{aligned}
\label{eq:jcas_model}
\end{equation}
where $x \in \mathbb{C}$ is the transmit signal from a base station, $y \in \mathbb{C}$ is the downlink ouput and $z\in \mathbb{C}^{N_R}$ is the received echo signal reflected from a target. $n_1$ and each element of $n_2$ are independent complex Gaussian noise with zero mean and unit variance. The channel state to be estimated is the angle of arrival $\phi$, and $a(\phi)$ is in the form of the steering vector of a uniform linear array with $N_R$ antenna elements
\begin{equation}
    a(\phi) = [1, e^{j \pi \sin \phi}, \cdots, e^{j\left(N_{R}-1\right) \pi \sin \phi}]^\top.
\end{equation}
The distortion is measured by the squared error $d(s,\hat{s}) = (s - \hat{s})^2$ and the input cost is constrained by $\int  \|x\|_2^2  \mu(\dx) \le P$. $\phi$ is assumed to be uniformly distributed in the range $[-\frac{\pi}{2}, \frac{\pi}{2}]$. Note that the optimal estimator in this case takes the form of \ac{mmse} estimator, given by the \ac{pm}
\begin{equation}\label{eq:h_Q}
    h^*_M(x,z) \triangleq \int \phi\, p_{\Phi|XZ}(\phi|x,z) \mathrm{d}\phi,
\end{equation}
and can be approximated similarly to \eqref{eq:p_s__xz2} as
\begin{equation}\label{eq:h_cm}
\begin{split}
    h^*_M(x,z) \approx \frac{1}{N_s} \sum_{j=1}^{N_s} \frac{\phi_j \cdot p_{Z|X\Phi}(z|x,\phi_j)}{\frac{1}{N_s}\sum_{i=1}^{N_s} p_{Z|X\Phi}(z|x,\phi_i)}.
\end{split}
\end{equation}
With the proposed method, we also design a \ac{nn} to parameterize the state estimator as $h_\theta$. The \ac{nn} structure is a simple fully-connected network with three layers and ReLU as the activation function. The real and imaginary parts of the input $(x,z)$ are concatenated to form a real-valued vector.

% As for the design of the estimator $h$, we design a NN to obtain the estimated angle $\hat{s}$ with learning rate 0.001, which can be written as
%   $ \hat{s} = W_3 \cdot ReLU(W_2 \cdot ReLU(W_1 \cdot [x; z] + b_1) + b_2) + b_3$. At a special case, we can get MMSE estimator by combining  \eqref{eq:h_Q} and importance sampling technique:

% \begin{equation}\label{eq:h_cm}
% \begin{split}
%     h^*_Q(x,z) \approx& \frac{1}{N} \sum_{j=1}^N \frac{s_j \cdot \hat{p}_{S|XZ}(s_j|x,z) }{p_S(s_j)}\\
%     =& \frac{1}{N} \sum_{j=1}^N \frac{s_j \cdot p_{Z|XS}(z|x,s_j)}{\frac{1}{N}\sum_{i=1}^N p_{Z|XS}(z|x,s_i)}.
% \end{split}
% \end{equation}

\begin{figure}
    \centering
    \includegraphics[width=0.48\textwidth]{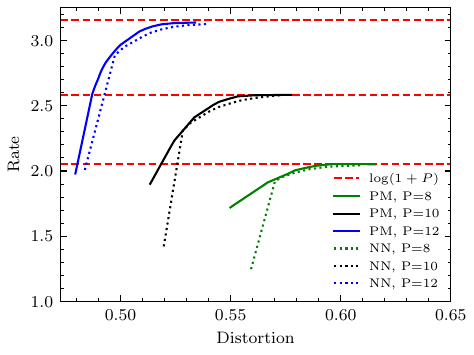}
    \caption{Computed \ac{cdc} curves for different input power $P$.}
    \label{fig:CD function}
\end{figure}

\begin{figure}
    \centering
    \includegraphics[width=0.48\textwidth]{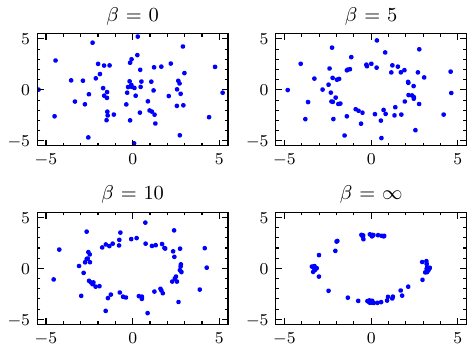}
    \caption{Constellation of particles in `PM' case with $P=10$dB.}
    \label{fig:particles_pm}
\end{figure}

\begin{figure}
    \centering
    \includegraphics[width=0.48\textwidth]{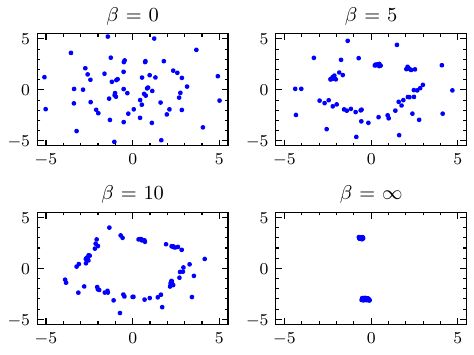}
    \caption{Constellation of particles in `NN' case with $P=10$dB.}
    \label{fig:particles_nn}
\end{figure}

We perform Algorithm~\ref{alg:compute} for both $h^*_M$ and $h_\theta$, corresponding to `PM' and `NN', respectively, with $N_R=2$, $N = 128$, $N_s = 128$, $N_y = 64$, $N_z = 8$ and $10 \log_{10} P$ taking $8$dB, $10$dB and $12$dB. After convergence, we test the input particles and estimator with $N_s = 128$, $N_y = 1024$, $N_z = 1024$. The resulting rate-distortion curve is shown in Fig. \ref{fig:CD function}, and the constellations of resulting
input particles are presented in Fig.~\ref{fig:particles_pm} and Fig.~\ref{fig:particles_nn}. The red dashed lines in Fig. \ref{fig:CD function} correspond to the case of $\beta=0$, i.e., no state estimation task is performed, in which the channel capacity takes the form of $\log(1+P)$, and the optimal input is Gaussian distributed with zero mean and variance $P$, as validated by the proposed method. As $\beta$ increases, the computed rate and distortion both decrease as expected, showing a concave and non-decreasing curve, which is studied in\cite{ahmadipour2022information}. Moreover, the input particles tend to concentrate on certain fixed points when $\beta \to \infty$, as illustrated in Fig.~\ref{fig:particles_pm} and Fig.~\ref{fig:particles_nn}, showing the random-deterministic trade-off in \ac{isac} systems. Interestingly, the concentrating particle locations of `NN' case are only two points compared to the four points obtained by `PM' case, which might result from the limited expression power of the chosen simple \ac{nn} structure.

\section{Conclusion}
This paper presents a numerical method to compute the \ac{cdc} function over continuous memoryless channels. Besides the objective values, the algorithm is also capable of obtaining the empirical input distribution and state estimation function, which is practically helpful in designing waveforms and estimators for better performance. For future works, we plan to analyze the convergence behavior of the alternating update steps and the overall computational complexity of the proposed method.

\bibliographystyle{IEEEtran}
\bibliography{IEEEabrv,mybib.bib}
\end{document}